\documentclass[a4paper,11pt]{article}
\pdfoutput=1 %

\usepackage{jcappub} %

\usepackage[T1]{fontenc} %

\usepackage[utf8]{inputenc}
\usepackage{aas_macros}
\usepackage{amsfonts}
\usepackage{amsmath}
\usepackage{amssymb}
\usepackage{graphicx}
\usepackage[dvipsnames]{xcolor}
\usepackage{hyperref}
\usepackage{soul}
\hypersetup{colorlinks=true,allcolors=teal}

\newcommand{\Msun}{\ensuremath{M_{\odot}}}
\newcommand{\Mh}{\ensuremath{h^{-1}M_{\odot}}}

\newcommand{\Mpch}{\ensuremath{h^{-1}{\rm Mpc}}}

\newcommand{\avg}[1]{\ensuremath{\left\langle \,#1\, \right\rangle}}

\newcommand{\figref}[1]{figure~\ref{#1}}

\newcommand{\secref}[1]{section~\ref{#1}}

\newcommand{\be}{\begin{equation}}
\newcommand{\ee}{\end{equation}}

\newcommand{\conditionalincludegraphics}[3][]{ \IfFileExists{#2}{\includegraphics[#1]{#2}}{\includegraphics[#1]{#3}} }

\title{\boldmath Dynamics of the response of dark matter halo to galaxy evolution in IllustrisTNG}

\author{Premvijay Velmani}
\author{and Aseem Paranjape}
\affiliation{Inter-University Centre for Astronomy \& Astrophysics,\\ Ganeshkhind, Post Bag 4, Pune 411007, India}

\emailAdd{premv@iucaa.in}
\emailAdd{aseem@iucaa.in}

\abstract{
We present the dynamical evolution of the dark matter's relaxation response to galaxies and their connection to the astrophysical properties as simulated in the IllustrisTNG suite of cosmological hydrodynamical simulations.
Our results show that the radially-dependent linear relaxation relation model from our previous work is applicable at least from redshift \(z=5\). We focus on the offset parameter \(q_0\), which characterizes the relaxation of dark matter shells without changing the enclosed mass.
We perform multiple time-series analyses to determine the possible causal connections between the relaxation mechanism and astrophysical processes such as star formation and associated feedback processes, as well as feedback due to active galactic nuclei.
We show that star formation activity significantly influences the halo relaxation response throughout its evolutionary history, with essentially immediate effects in the inner haloes and delayed effects of 2 to 3 Gyr in the outer regions. Metal content shows a weaker connection to relaxation than star formation rates, but the accumulated wind from feedback processes exhibits a stronger correlation.
These findings enhance our understanding of halo relaxation mechanisms. 
Our estimates of the time-scales relevant for dark matter relaxation can potentially improve the description of halo profiles in existing baryonification schemes and semi-analytical galaxy formation models. Our results also show how
the relaxation response of dark haloes can probe the evolutionary history of the galaxies they host.}

\keywords{dark matter haloes, galaxies, cosmology:simulations, adiabatic relaxation, halo response}

\begin{document}
\maketitle
\flushbottom

\section{Introduction}
\label{sec:intro}

In the standard paradigm of cosmology, galaxies form within the gravitationally collapsed structures called haloes, primarily made of dark matter that doesn't interact with baryonic galaxies except through gravity \citep[][]{wr78}. Formation and the evolution of these haloes and the galaxies they host are crucial in both cosmological and astrophysical studies. %
In the $\Lambda$CDM paradigm, dark matter haloes form through the gravitational collapse of initial density fluctuations \citep{1974ApJ...187..425P,2002PhR...372....1C}. Properties of these haloes, such as their triaxial shapes \citep{1988ApJ...327..507F} and universal mass profiles (NFW profiles; \citep{1996ApJ...462..563N,1997ApJ...490..493N}), have been well-characterized in gravity-only simulations. However, the presence of baryonic matter introduces additional complexities. The gravitational coupling between baryons and dark matter can significantly alter the spatial distribution and evolution of the latter, necessitating a detailed study of this interaction. 

High-resolution cosmological hydrodynamical simulations, which incorporate detailed feedback processes from supernovae and active galactic nuclei (AGN), provide a more nuanced understanding of halo dynamics (e.g., OWLS \citep{2010MNRAS.402.1536S}; Illustris \citealp{2014MNRAS.445..175G}; FIRE \citep{2014MNRAS.445..581H}; EAGLE \citep{2015MNRAS.446..521S}). These simulations reveal that feedback mechanisms can significantly alter the inner density profiles of dark matter haloes, potentially resolving the discrepancy between observed dark matter cores and the cuspy profiles predicted by gravity-only simulations \citep{2014Natur.506..171P}. However, the degree of transformation depends on factors such as the time-scale and frequency of feedback events \citep{2012MNRAS.421.3464P,2014ApJ...793...46O}.

The response of dark matter haloes to galaxy formation includes two main aspects: contraction or expansion towards the centre and changes in their triaxial shape \citep[][]{2004ApJ...616...16G,2006PhRvD..74l3522G,2010MNRAS.402..776P,2010MNRAS.406..922T,2010MNRAS.405.2161D,2010MNRAS.407..435A,2011MNRAS.414..195T,2016MNRAS.461.2658D,2019A&A...622A.197A,2022MNRAS.511.3910F,2023Velmani&Paranjape}. 
In this study, we focus on the former aspect, examining spherically averaged mass profiles.
In a previous work \cite{2023Velmani&Paranjape}, we examined this impact of baryonic processes on dark matter haloes, focusing on the radial halo profiles in IllustrisTNG and EAGLE simulations across a wide range of halo masses at redshift $z \sim 0$. We found that simple adiabatic contraction models, which assume spherical symmetry and no shell crossing, often fall short of accurately describing the complex response of dark matter to baryonic effects observed in high-resolution hydrodynamical simulations.

However, by empirical modelling within the context of quasi-adiabatic relaxation, simple relations were found to describe the nature of the halo relaxation response. 
Building on our previous study, the present work aims to systematically investigate the dynamical evolution of dark matter halo relaxation and its connection to the evolution of halo and galaxy properties. In this paper, we use data from the IllustrisTNG simulation project towards this goal. By comparing the properties of haloes in the baryonic simulations of this suite with their counterparts in the corresponding gravity-only simulations, we characterize the relaxation behaviour of a variety of haloes across a wide range of masses over a time from redshift $z \sim 5$ to $z \sim 1$.

The structure of this paper is as follows.
We start with a brief description of the numerical simulations and the construction of a catalogue of matched haloes along with their evolutionary tracks in \secref{sec:simhals}. Then, in \secref{sec:methods-relchar}, we characterize the relaxation response in the context of a quasi-adiabatic framework and study their evolution in halo populations. This is followed in \secref{sec:methods-stat} by a statistical exploration of the connection between halo properties such as star formation rate and metallicity and their role in mediating the halo relaxation response. This analysis especially focuses on the rank correlations between these quantities not only across haloes in the populations but also over time. %
Finally, we conclude with a summary of our key findings and their applications in \secref{sec:conclusion}. Throughout this paper, $\ln$ and $\log$ denote the natural and base-10 logarithms, respectively.

\section{Simulations and haloes}
\label{sec:simhals}
In this work, we use all three different cosmological volumes of the publicly available IllustrisTNG simulations at their highest resolution \citep{2019ComAC...6....2N}; details of these simulations are outlined in \secref{sec:methods-itng}. Then, we describe in \secref{sec:methods-halopairsel} the identification and selection of haloes across a wide range of masses along with their corresponding matched halo simulated without any baryonic astrophysics. We need evolving haloes to study the dynamical relaxation; these are constructed using the merger trees consisting of the progenitor haloes matched across time as described in \secref{sec:methods-tracehals}. Finally, in \secref{sec:hal-gal-props}, %
we explore some key properties of the halo (and its central galaxy) and their overall evolution in the population. 

\subsection{IllustrisTNG simulations}
\label{sec:methods-itng}
IllustrisTNG provides hydrodynamical simulations of three different cosmological volumes namely TNG50, TNG100, and TNG300, with periodic box sizes of $35 \Mpch$, $75 \Mpch$, and $200 \Mpch$, respectively, consistent with the cosmology from Planck collaboration \citep[][]{2016A&A...594A..13P}. In each box, we utilise the highest resolution runs with dark matter mass resolution of $5.4 \times 10^5 \Msun$, $8.8 \times 10^{6} \Msun$ and $7.0 \times 10^7 \Msun$ respectively from the smallest TNG50 to largest TNG300; along with their corresponding gravity-only simulation of same volumes. Initial conditions for these cosmological boxes were constructed using the \textsc{N-GenIC} code \citep[][]{2015ascl.soft02003S} with the Zel'dovich approximation \citep[][]{1970A&A.....5...84Z} at $z = 127$. All these simulations employed the \textsc{arepo} code \citep[][]{2020ApJS..248...32W} utilizing a moving mesh approach defined by Voronoi tessellation \citep[][]{2010MNRAS.401..791S} for the hydrodynamics. These simulations incorporate a state-of-the-art subgrid prescription for the major baryonic processes such as cooling, star formation, stellar, and AGN feedback along with cosmic magnetic field \citep[][]{2017MNRAS.465.3291W,2018MNRAS.473.4077P}. In this work, we utilise data from redshift $z=0$ all the way to redshift $z=5$ for both hydrodynamical and corresponding gravity-only runs.

\subsection{Halo selection}
\label{sec:methods-halopairsel}
Haloes in these simulations were identified using a friend-of-friends (FoF) algorithm \citep[see][for specifics]{2019ComAC...6....2N}, and the gravitationally bound substructures within these FoF group haloes were found using \textsc{subfind} code \citep{2001MNRAS.328..726S} and identified as subhaloes. While there can be more than one subhalo within a given FoF halo, only the central subhalo encloses the halo centre, which is defined by the co-moving position of the minimum gravitational potential. The radius of the sphere around this centre enclosing a mean matter density that is 200 times the cosmological critical density is defined as the `virial' radius $R_{\rm vir}\equiv R_{\rm 200c}$ of a given FoF group halo; while the total mass enclosed within this radius quantifies the mass of the halo $M\equiv M_{200c}$.

Following our previous work \citep{2023Velmani&Paranjape}, we match the haloes from the full hydrodynamic simulations with the haloes in the corresponding gravity-only runs performed with the same seed for initial conditions. This gives a set of matched halo pairs showing the strongest overlap in their proto-haloes among the nearby haloes of similar sizes found by the KD-tree algorithm. From this catalogue, we select populations of matched halo pairs by the logarithmic mass of the gravity-only halo $(\log(M/\Mh))$ in bins centred at $11.5, 12, 12.5, 13, 13.5, 14$ with a bin width of 0.3 at redshift $z=0.01$. While the small volume TNG50 offers well-resolved low-mass haloes $10^{11.5} \Mh$ and $10^{12} \Mh$, the TNG100 cosmological box gives $10^{12.5} \Mh$ and $10^{13} \Mh$ haloes and the largest volume TNG300 provides an adequate number of cluster-scale haloes of masses $10^{13.5} \Mh$ and $10^{14} \Mh$.

\subsection{Tracing evolutionary history}
\label{sec:methods-tracehals}
In cosmological simulations performed with N-body techniques for the dark matter,  it is particles that are evolved in cosmological volumes, and hence haloes need not be tracked explicitly. Rather, the haloes found at different snapshots in time are matched to construct the evolutionary tracks of the haloes. In this work, we identify the central subhaloes pairs corresponding to each pair of FoF group halos at $z\sim 0$ in our catalogue. These central subhaloes are traced back along the most massive progenitor branch using merger trees constructed by the \textsc{SubLink} code \citep{2015RodriguezGeneletalSubLink}. Then, the corresponding host FoF group haloes are considered as the progenitors of the matched haloes in our catalogue at $z \sim 0$. This gives us a catalogue of evolving pairs of haloes from redshift $z \sim 5$ to $z \sim 0$ corresponding to a cosmological look-back time of $12.63$ Gyr.

From this catalogue, we remove all those haloes that had a major merger during this period with a dark matter mass ratio above four. Further, we also remove those pairs of evolving haloes that are either separated by more than their virial radius or have sizes differing by more than $25\% $ at any given redshift. This leaves us with a sample of evolving matched haloes for around $50\%$ to $80\%$ of the populations of all haloes in our $z \sim 0$ catalogue. For these halo pairs, we study the relaxation by comparing the spherically averaged radial distribution of matter within their virial radii; this is described in  \secref{sec:methods-relchar}. Before that, we explore some of the key properties of these haloes in \secref{sec:hal-gal-props} that are relevant in this work.

\subsection{Halo and galaxy properties}
\label{sec:hal-gal-props}
\begin{figure}[htbp]
\centering
\includegraphics[width=\linewidth]{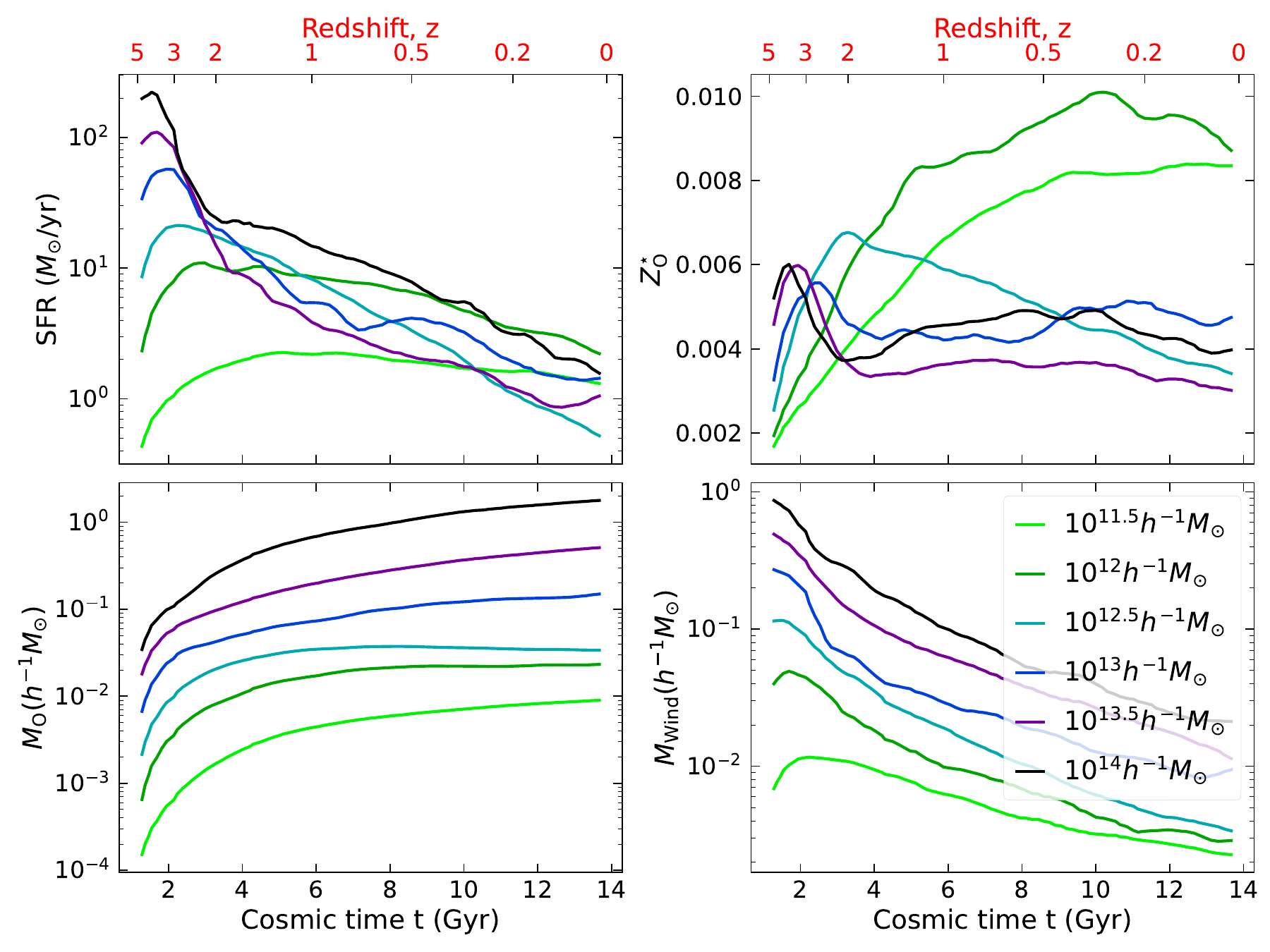}
\caption{Evolution history of the average star formation rate (SFR, top left panel), mean oxygen metallicity in star-forming regions ($Z^{\star}_{\rm{O}}$, top right panel), the mass of oxygen in gas ($M_{\rm{O}}$, bottom left panel) and mass of wind ($M_{\rm{Wind}}$, bottom right panel) in halo populations selected by their total mass at redshift $z\sim 0$ in the gravity-only run.}
\label{fig:evolution-hal-gal-props}
\end{figure}

When it comes to the role of galactic astrophysical processes in mediating the halo relaxation, feedback processes are known to play a significant role \cite{2011MNRAS.414..195T,2023Velmani&Paranjape}. To understand this, we look into the properties of the halo that quantify various galactic processes involved in producing the feedback. 
In this work, we primarily focus on the  following quantities: 
\begin{itemize}
    \item \textbf{SFR}: The net star formation rate of the central subhalo. The feedback associated with stellar winds and supernovae of massive stars follows star formation activity.
    \item $\mathbf{Z^{\star}_{\rm{O}}}$: The mean metallicity of oxygen in the star-forming regions of the central subhalo within twice the stellar half mass radius. Oxygen is a key tracer of chemical enrichment and feedback processes.
    \item $\mathbf{M_{\rm{Wind}}}$: The total mass of wind within the halo group. In addition to the stellar feedback, accretion onto AGN can also produce powerful winds. These winds eventually lose their mass and energy as they travel through the gas.
    \item $\mathbf{M_{\rm{O}}}$: The total oxygen content within the halo group in the gas component; This quantity probes the stellar feedback over a longer time, even after the suppression of $\mathbf{Z^{\star}_{\rm{O}}}$ due to inflowing metal-poor gas.
\end{itemize}
All these quantities are defined for the hydrodynamic halo of each matched group halo pair in our catalogue. The evolution history of these quantities in the populations of haloes having the same final mass is shown in \figref{fig:evolution-hal-gal-props}. While low mass haloes at $z \sim 0$ are currently at their peak star formation, it is significantly suppressed in the high mass haloes.

\section{Characterizing Relaxation Response}
\label{sec:methods-relchar}
The overall relaxation response of dark matter in a halo, such as an expansion or contraction in response to galaxy formation, is usually analysed through changes in the spherically averaged mass profiles. In our catalogue of matched evolving pairs of haloes, while the hydrodynamic ones with galaxies include this relaxation response, the gravity-only counterparts provide the corresponding unrelaxed dark haloes. These radial dark matter mass profiles are obtained as the cumulative sum of the mass contributed by all dark matter particles within concentric spherical shells. In the case of the gravity-only halo, we consider the cosmic dark matter fraction of the mass in each particle of the gravity-only halo to be contributing to the dark matter. Characterizing the relaxation from these profiles is done using the quasi-adiabatic relaxation framework as described below.

\subsection{Quasi-adiabatic relaxation framework}
\label{sec:methods-adiab}
All the relaxation response on cold dark matter happens entirely through gravitational encounters with the baryons. In particular, it is an integrated effect of the flow of baryonic mass in the past due to galactic processes such as inflows and feedback. The quasi-adiabatic relaxation is a physically motivated framework to model the change in the spherically averaged dark matter distribution at a given time as a function of just the spherically averaged baryonic distribution at that same time. This baryonic profile includes all the mass other than the dark matter, including the mass in gas cells assigned by a Gaussian kernel to concentric spherical shells. 

Early works made further assumptions that the halo is spherical and that the dark particles maintain their radial ordering while responding adiabatically to the flow of the baryonic particles across them \citep[][]{1986ApJ...301...27B}. Suppose a dark matter particle at radius $r_i$ in the unrelaxed
halo ends up in radius $r_f$ in the relaxed halo, then the enclosed dark matter within spheres of those radii will be equal.
\be 
M_f^d(r_f) = M_i^d(r_i)\,.
\label{eq:DMmass}
\ee
However, the \emph{total} mass enclosed within those spheres is not necessarily equal $M_i(r_i) \neq M_f(r_f)$ due to the flow of baryonic mass. When the dark matter particles conserve angular momentum, maintaining nearly circular orbits, this change in total mass enclosed must be consistent with the amount of relaxation,
\citep[][]{1986ApJ...301...27B},
\begin{align}
    r_i \,M_i(r_i) = r_f \,M_f(r_f) %
    \implies 
\frac{r_f}{r_i} = \frac{M_i(r_i)}{M_f(r_f)}\,. 
\label{eq:AR}
\end{align}
The quasi-adiabatic relaxation framework is an empirical extension to this idealised scenario and considers the relaxation ratio $r_f/r_i$ as a function of the mass ratio $M_i/M_f$.
\begin{align}
\frac{r_f}{r_i} &= 1 + \chi \left( \frac{M_i(r_i)}{M_f(r_f)} \right) 
\label{eq:qAR}
\end{align}
In a simple extension, the baryonification procedures in \cite{2015JCAP...12..049S,2021MNRAS.503.4147P} include dark matter response as a quasi-adiabatic relaxation with $\chi(y) = q (y-1)$. However, a variety of quasi-adiabatic models have been proposed \citep{2004ApJ...616...16G,2010MNRAS.407..435A,2023Velmani&Paranjape}.  For example, to include the effects of non-circular orbits of dark matter particles, Gnedin et al. (2004) \citep{2004ApJ...616...16G} provided a empirical model with parameters calibrated by simulation data. However, we have checked that many of those models were still significantly different from the relaxation response of a wide variety of haloes produced by IllustrisTNG, atleast with the given parameter values \citep[see appendix of ][]{2023Velmani&Paranjape}. Instead, we found that including a explicit dependence on the halo-centric distance gave a more universal description across a variety of haloes in IllustrisTNG \cite{2023Velmani&Paranjape}. And to account for the effects of astrophysical feedbacks, we have introduced the relaxation offset, a quantity that was found to significantly correlate with some of the halo/galaxy properties at $z=0$ \citep{2023Velmani&Paranjape}. We primarily study this quantity here to further understand the dynamical connection between relaxation response and the astrophysical processes; This quantity is described below.

\subsection{Relaxation offset}
In our previous work \cite{2023Velmani&Paranjape}, we found that 
the relaxation relation follows a locally linear relation. 
\begin{align}
    \label{eq:chi-linear-q0}
    \frac{r_f}{r_i} - 1 &= q_1(r_f) \left[ \frac{M_i(r_i)}{M_f(r_f)} - 1 \right] + q_0(r_f)\,.
\end{align}
For a given halo sample, at each $r_f$, the relation between mass ratio and relaxation ratio across all the haloes is fitted by a linear curve to obtain the parameters $q_0(r_f)$ and $q_1(r_f)$. It was also found that $q_0(r_f)$ is usually roughly uniform from the inner to the outer haloes. In this work, we have tested that this radial-dependent linear model is a good description of the halo relaxation over their evolutionary history for all six halo masses considered. Akaike weights, estimated from the corrected Akaike Information Criterion (AICc) \cite{1974Akaike,1978Suguira}\cite[see][for a review]{2007Liddle}, show that the first-order polynomial is a better fit for the relaxation relations along the most-massive progenitor branch to redshift $z=5$ and at halo-centric distances ranging from the virial radius to at least $5\%$ of that radius.\footnote{At a given halo mass, redshift, and halo-centric distance, we group the haloes into bins of mass ratios ($M_i/M_f$) and then fit the mean $r_f/r_i$ as a function of $M_i/M_f$, weighted by the inverse error. This provides the AICc values and the Akaike weights that quantify the contribution of each model in the Bayesian average predictions. We find that the first-order polynomial dominates over the zeroth- and second-order polynomials across the halo-centric distances and redshifts considered for each of the halo masses.} To characterize the halo-to-halo variation in the relaxation offset, we define the following relaxation quantity for each halo in the population. 
\begin{align}
\label{eq:def-q0hal}
q_0 |_{h}(r_f) &\equiv \frac{r_f}{r_i} - 1 - q_1(r_f) \left[ \frac{M_i(r_i)}{M_f(r_f)} - 1 \right]\,.
\end{align}
This quantity $q_0 |_{h}(r_f)$ is usually uniform within each of the haloes; We primarily focus on its mean in this work, denoted simply as $q_0$ for that halo. We call this a relaxation offset parameter since it characterizes the amount of relaxation of dark matter shells that have no change in the total enclosed mass. In addition to this, we also study the mean $q_0(r)$ in the inner ($\sim 10\%~R_{\rm{vir}}$) and the outer ($\sim 50\%~R_{\rm{vir}}$) halo defined as $q_0^{\rm{in}} ~\& ~q_0^{\rm{out}}$ respectively.

\begin{figure}[htbp]
\centering
\includegraphics[width=\linewidth]{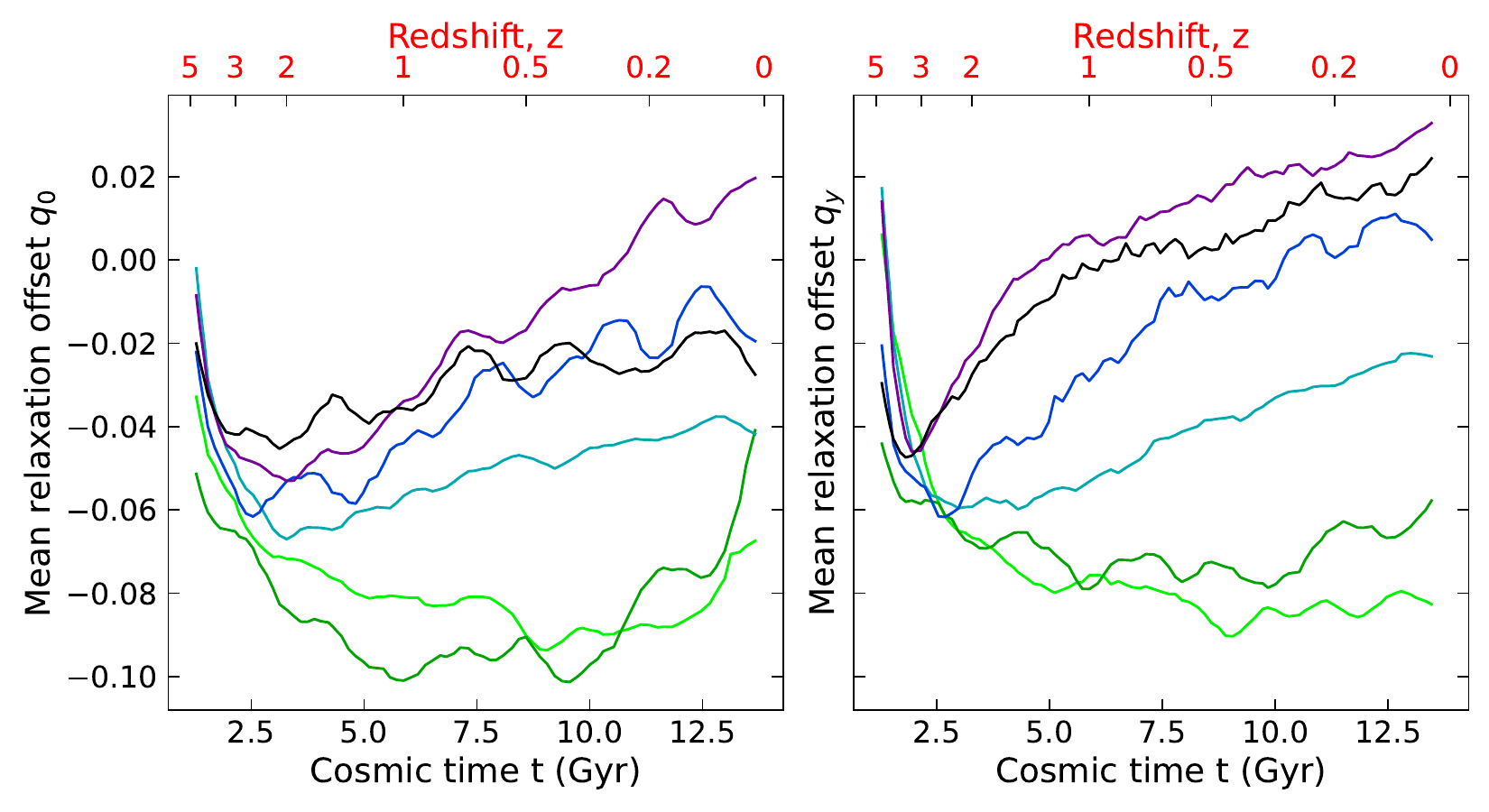}    \caption{Dynamical evolution of the relaxation offset parameters $q_0$ (left panel) and $q_y$ (right panel) averaged in halo populations selected by their mass indicated by the same colour coding as in \figref{fig:evolution-hal-gal-props}}
\label{fig:evolution-hal-reln-offset}
\end{figure}
In addition to this model parameter $q_0$, we also define a somewhat model-agnostic quantity to characterize just the offset in the relaxation. For each matched halo, the y-intercept of the relaxation relation is given by the relation between $M_i/M_f-1$ and $r_f/r_i-1$ for each matched halo. This parameter, denoted as $q_y$, is the offset in the relaxation ratio $r_f/r_i$ from unity for the shells having a mass ratio of unity $M_i/M_f=1$. The evolution history of these relaxation offset parameters for our halo populations is shown in \figref{fig:evolution-hal-reln-offset}. We find that the average evolution of $q_0$ and $q_y$ are similar for all haloes. The relaxation offset starts with smaller values and reaches peak magnitudes before approaching zero, consistent with the $z \sim 0$ results from our previous work \citep{2023Velmani&Paranjape}. We have also checked that they show good correlation across haloes, especially $q_y$ shows good correlation with the $q_0^{\rm{in}}$. Hence, in this study, we will primarily focus the relaxation offset parameters $q_0(r)$,  $q_0^{\rm{in}} ~\& ~q_0^{\rm{out}}$.

\section{Statistical analysis}
\label{sec:methods-stat}
The average evolution of the relaxation parameters shown in \figref{fig:evolution-hal-reln-offset} and other properties shown in \figref{fig:evolution-hal-gal-props} already suggest that the relaxation offset is strongest following the peak star formation period. However, we must study their evolution in individual haloes to further understand the role of different astrophysical processes. This additional information can be captured using correlations not only across haloes in the population but also over time between these quantities. In the \secref{sec:sfr-metallicity} below, we develop this methodology as we explore the connection between star formation rate and metallicity ($Z^{\star}_{\rm{O}}$) through these correlations. In the following (sub)section, we use this method to study the connection between relaxation and various galactic processes. 

\begin{figure}[bhtp]
\centering
\includegraphics[width=\linewidth]{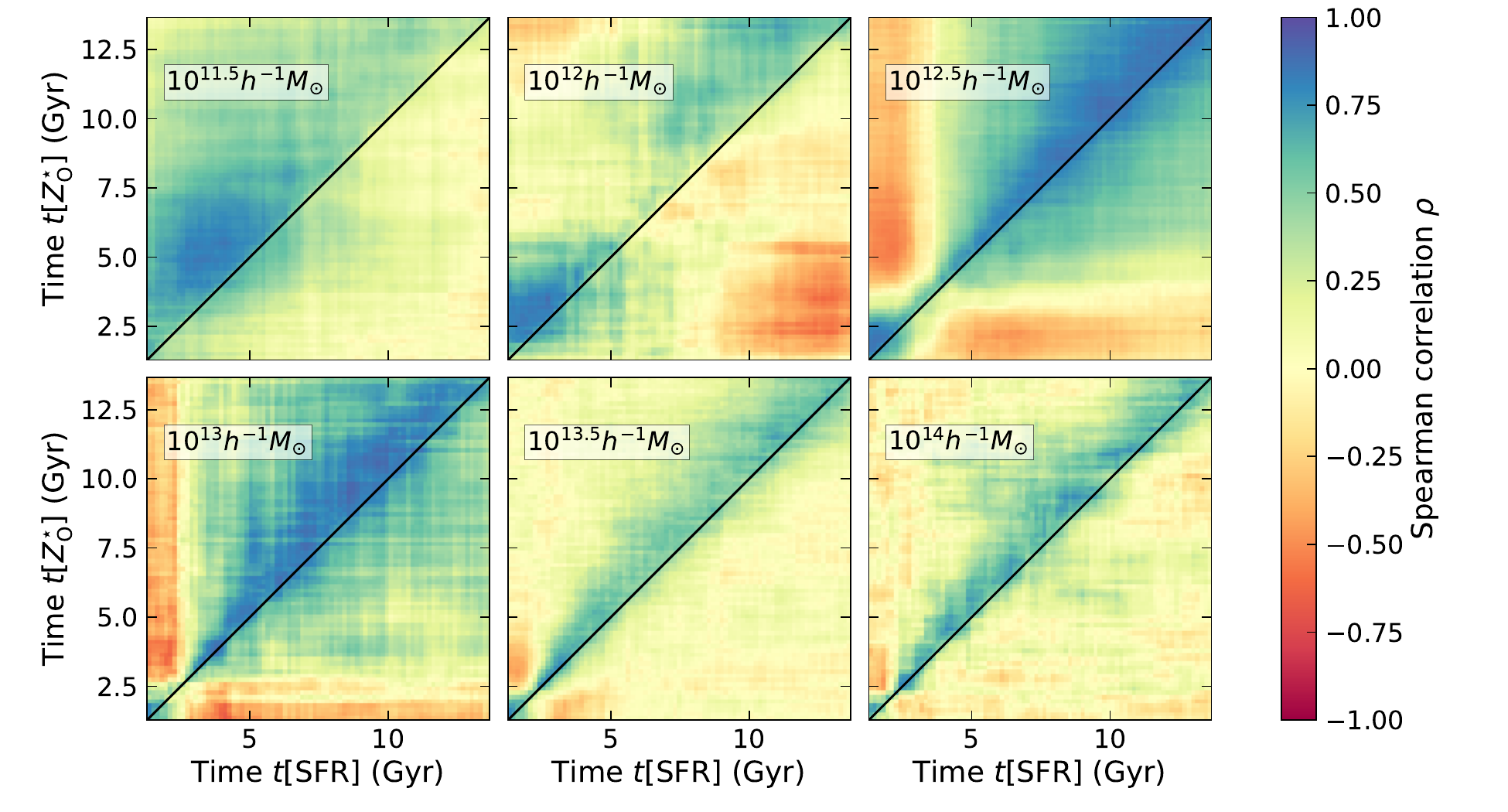}
\caption{Correlation across haloes between star formation rate and metallicity ($Z^{\star}_{\rm{O}}$) in halo populations selected by their final masses at redshift $z\sim 0$. In the images, the colour at a given $(t[\rm{SFR}],t[Z^{\star}_{\rm{O}}])$ represents the Spearman rank correlation coefficient between the SFR at time $t[\rm{SFR}]$ and metallicity at time $t[Z^{\star}_{\rm{O}}]$ across haloes for each pair of times.}
\label{fig:dynam-correl-sfr-ZOsfr-img}
\end{figure}

\subsection{Star formation rate and metallicity connection}
\label{sec:sfr-metallicity}
As a warm-up, and as a means to showcase our method, we first present the correlation between the $Z^{\star}_{\rm{O}}$ 
and the SFR for haloes selected by their mass at redshift $z\sim 0$. 
In the simulation, metals are added to the gas by various feedback processes such as stellar winds and supernovae explosions that are dominantly produced by the short-lived massive stars. Hence, the metallicity is expected to rise following periods of star formation activity. To characterize this information, we focus on the Spearman rank correlations between the SFR and $Z^{\star}_{\rm{O}}$ at different times.

For a given population of $N$ haloes, we have the values of these two quantities for individual haloes tracked in distinct time steps $t_i$ with $i=1,2,\ldots,T$ separated by a uniform time interval $\Delta t = 157$ Myr. Using this, we construct the $T \times T$ matrix of Spearman correlation coefficients denoted by $\rho^s_{ij} [ \rm{SFR}, Z^{\star}_{\rm{O}}]$. Each element $ij$ of this matrix corresponds to the correlation between SFR at time $t_i$ with $Z^{\star}_{\rm{O}}$ at time $t_j$. (Note that the matrix is not expected to be symmetric.) Suppose $\rm{SFR}(n,t_i)$ and $Z^{\star}_{\rm{O}}(n,t_i)$ denote the corresponding quantities in the $n^{\rm{th}}$ halo at time $t_i$ in a population of $N$ haloes, then,
\begin{align}
\rho^s_{ij} [\rm{SFR}, Z^{\star}_{\rm{O}}] &= \rho^s \left[~\rm{SFR}(\{1 \ldots N\},t_i), ~Z^{\star}_{\rm{O}}(\{1 \ldots N\},t_j)~\right] \\
\intertext{where the Spearman correlation $\rho^s$ is defined as the Pearson correlation ($\rho$) of the ranks ($R$), which in turn can be expressed as the covariance normalized by their standard deviations ($\sigma$).}
\rho^s_{ij} [\rm{SFR}, Z^{\star}_{\rm{O}}] &= \rho \left[~R[~{\rm{SFR}}(\{1 \ldots N\},t_i)~], ~R[~Z^{\star}_{\rm{O}}(\{1 \ldots N\},t_j)~]~\right]\\
&= \frac{\text{cov}\left(R[~{\rm{SFR}}(\{1 \ldots N\},t_i)~], R[~Z^{\star}_{\rm{O}}(\{1 \ldots N\},t_j)~]~\right)}{\sigma\left(R[{\rm{SFR}}(\{1 \ldots N\},t_i)]~\right) \times \sigma\left(R[Z^{\star}_{\rm{O}}(\{1 \ldots N\},t_j)]~\right)}\\
&= \frac{  (1/N) \sum_{n=1}^{N} R_{\rm{SFR}}(n,t_i) \times  R_{Z^{\star}_{\rm{O}}}(n,t_j) - \mu_{n} [ R_{\rm{SFR}}(n,t_i) ]  \times \mu_{n} [R_{Z^{\star}_{\rm{O}}}(n,t_j)] }
{ \sigma_{n} [ R_{\rm{SFR}}(n,t_i) ]  \times \sigma_{n} [R_{Z^{\star}_{\rm{O}}}(n,t_j)] }
\end{align}
where $R_{\rm{SFR}}$ and $R_{Z^{\star}_{\rm{O}}}$ denote the corresponding ranks, while $\mu_n$ and $\sigma_n$ represent the mean and standard deviation, respectively computed across haloes ($n=1$ to $N$) at each time step. This matrix is depicted in \figref{fig:dynam-correl-sfr-ZOsfr-img} for the six populations of our evolving haloes selected by their final masses at $z \sim 0$; We find blue patches of strong positive correlation mostly near the diagonal. 
More specifically, there is usually a stronger positive correlation between the star formation rate at a given time and the $Z^{\star}_{\rm{O}}$ in the near future time. This is expected since the metals are primarily produced by the feedback that follows the star formation activity. 

Another thing to note is the strong negative correlation between the star formation rate of the haloes at later times with metallicity at a very early time and vice versa. 
This happens at a characteristic time scale associated with the peak star formation period for that population of haloes. 
For example, in $10^{13} \Mh$ haloes, this transition period happens at a cosmic time of $2$ Gyr; this is indicated by the transition between blue and red patches. 
While the metallicity is initially raised by the feedback events, the same feedback pushes inner metal-rich gas into the circumgalactic medium (CGM). On the other hand, the inflowing metal-poor gas from CGM lowers the metallicity while being crucial to feed the star formation activity at a later time. 
This leads to the inverse relationship between star formation rate and metallicity at fixed stellar mass reported in the literature \citep[][]{2011DaveFinlatorOppenheimer,2018TorreyVogelsberger_etal_SFRZ}; and our results are consistent with it.

\subsubsection{Correlation at fixed time intervals}
\label{sec:halhal-corr-sfrZ}
\begin{figure}[htbp]
\centering
\includegraphics[width=.6\linewidth]{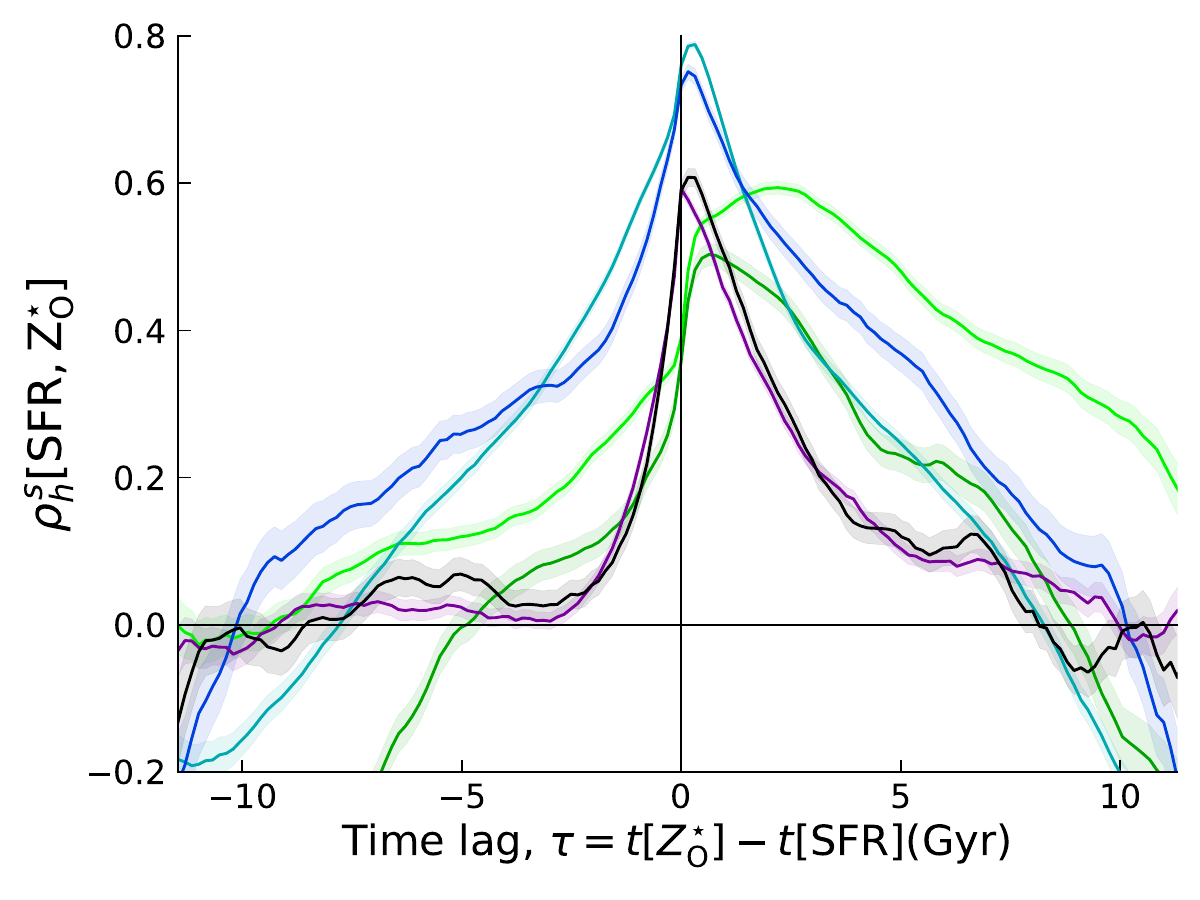}
\caption{Spearman correlation $\rho^s_h(\tau)$ across haloes between SFR and metallicity ($Z^{\star}_{\rm{O}}$) with a time lag. This is shown for the six populations of haloes selected by their final masses at redshift $z\sim 0$; Colour coding follows \figref{fig:evolution-hal-gal-props}.}
\label{fig:dynam-correl-sfr-ZOsfr-timeshift-func-halhal}
\end{figure}

We now focus on using the correlations to explore the repetitive nature of galactic processes. As a preliminary approach, we quantify this using the average of the Spearman rank correlation matrix $\rho^s_{ij}$ along the diagonals parallel to the black line shown in each panel of the \figref{fig:dynam-correl-sfr-ZOsfr-img}. This essentially represents the time-averaged correlation across haloes between SFR and $Z^{\star}_{\rm{O}}$ at a fixed earlier or later time, defined as:
\begin{align}
\rho^s_h(\tau)[\rm{SFR}, Z^{\star}_{\rm{O}}] &= \avg{ \rho^s [ \rm{SFR}(t), Z^{\star}_{\rm{O}}(t+\tau)] }_{t}  
\end{align}
This quantity is computed at distinct time steps $(\tau=m \Delta t)$ for integer $m = 0,\pm 1,\ldots,\pm T$ using the matrix $\rho^s_{ij}$ as,
\begin{align}
\label{eq:def-hal-correl}
\rho^s_h(\tau = m \Delta t) 
&= \avg{ \rho^s_{i,i+m} }_i \\
&= \begin{cases}
\sum_{i=1}^{T-m} \rho^s_{i,i+m}/(T-m) \qquad \text{if } m \geq 0 \\
\sum_{i=1-m}^{T} \rho^s_{i,i+m}/(T-m) \qquad \text{if } m < 0
\end{cases}
\end{align}
Similar to the standard Spearman correlation coefficient, this quantity takes on a value between $-1$ and $+1$ at each $\tau$, indicating the strength of the correlation from being strongly negative to positive.
As shown in \figref{fig:dynam-correl-sfr-ZOsfr-timeshift-func-halhal}, this analysis reveals that the positive correlation is strongest between the SFR and the immediate future $Z^{\star}_{\rm{O}}$ as expected. Also, notice that on average, there is a stronger positive correlation $\rho^s_h$ for $\tau>0$ than $\tau<0$; This is particularly evident for the lowest (green curves) and highest halo masses (black and purple curves). There is a relatively stronger positive correlation at intermediate masses (blue curves) indicated by higher peaks; However, it is much more symmetrical and the difference in the correlation of past and future metallicity with SFR is not clearly seen. This is a consequence of the fact that the $\rho^s_h(\tau)$ focuses only on the halo-halo correlation. Although $\rho^s_h(\tau)$ involves summation over both haloes and their evolutionary time steps, the ranks are computed independently at each of the time steps. When the halo-to-halo variations are much stronger than the temporal variations in the SFR and $Z^{\star}_{\rm{O}}$, then these ranks are less likely to change with time. Hence in such scenarios, $\rho^s_h(\tau)$ does not capture the correlation between temporal variations in the SFR and $Z^{\star}_{\rm{O}}$ of individual haloes. In the following section, we present another quantity that can directly probe the correlations over time.

\subsubsection{Time Correlations}
\label{sec:time-correl-sfrZ}
In order to further understand the connection between the sequences of galactic processes, we study time correlations that capture the temporal variations along the evolutionary tracks of each halo.
Consider an individual halo along with its evolutionary track, say $n^{\rm{th}}$ halo in a population of N haloes. We characterize the time correlation through the Spearman correlation as a function of time lag, $\rho^s_t |_{n}(\tau)[\rm{SFR}, Z^{\star}_{\rm{O}}]$ defined as follows.
\begin{align}
\rho^s_t |_{n}(\tau)[\rm{SFR}, Z^{\star}_{\rm{O}}] &= \rho^s [ ~\rm{SFR}(n,t), Z^{\star}_{\rm{O}}(n,t+\tau) ~] \quad \text{where } ~t, t+\tau \in [t_1,t_T].
\end{align}
This quantity is also computed at distinct time steps ($\tau = m \Delta t$) for integer $m=0, \pm1, \ldots, \pm T$ as:
\begin{align}
\label{eq:def-time-correl-nth-halo}
\rho^s_t |_{n}(\tau = m \Delta t) &= 
\begin{cases}
\rho^s [ ~\rm{SFR}(n,\{t_1 \ldots t_{T-m}\}), ~Z^{\star}_{\rm{O}}(n,\{t_{1+m} \ldots t_T\}) ~]  \qquad \text{if } m \geq 0 \\
\rho^s [ ~\rm{SFR}(n,\{t_{-m+1} \ldots t_T\}), ~Z^{\star}_{\rm{O}}(n,\{t_1 \ldots t_{T+m}\}) ~] \qquad \text{if } m < 0
\end{cases}.
\end{align}

Alternatively, cross-correlation is also a characterization of the correlation over time in that halo. In particular, consider the cross-correlation function between the normalized ranks of SFR and $Z^{\star}_{\rm{O}}$ over its evolutionary history given by, 
\begin{align}
\rho^s_{t'} |_{n}(\tau = m \Delta t) 
&= \begin{cases}
\sum_{i=1}^{T-m} R^{\rm{norm}}_{\rm{SFR}}(n,t_i) \times R^{\rm{norm}}_{Z^{\star}_{\rm{O}}}(n,t_{i+m})/(T-m) \qquad \text{if } m \geq 0 \\
\sum_{i=1-m}^{T} R^{\rm{norm}}_{\rm{SFR}}(n,t_i) \times R^{\rm{norm}}_{Z^{\star}_{\rm{O}}}(n,t_{i+m})/(T-m) \qquad \text{if } m < 0
\end{cases}
\end{align}
where the normalized ranks are defined as
\begin{align}
R^{\rm{norm}}_{\rm{SFR}}(n,\{t_1 \ldots t_T\}) &= \frac{R[\rm{SFR}(n,\{t_1 \ldots t_T\})] - \avg{R[\rm{SFR}(n,\{t_1 \ldots t_T\})]}}{\sigma[R[\rm{SFR}(n,\{t_1 \ldots t_T\})]]},\\
R^{\rm{norm}}_{Z^{\star}_{\rm{O}}}(n,\{t_1 \ldots t_T\}) &= \frac{R[Z^{\star}_{\rm{O}}(n,\{t_1 \ldots t_T\})] - \avg{R[Z^{\star}_{\rm{O}}(n,\{t_1 \ldots t_T\})]}}{\sigma[R[Z^{\star}_{\rm{O}}(n,\{t_1 \ldots t_T\})]]}.
\end{align}

Both $\rho^s_t$ and $\rho^s_{t'}$ probe the time correlations, but they are not necessarily equal. To illustrate the difference, consider the $n^{\rm{th}}$ halo's evolutionary history over a period of 12 billion years. For a time-lag of 3 billion years between SFR and $Z^{\star}_{\rm{O}}$, $\rho^s_{t} |_{n}(\tau=3~\rm{Gyr})[\rm{SFR}, Z^{\star}_{\rm{O}}]$ is the Spearman correlation coefficient between SFR data from 12 billion years ago to 3 billion years ago and the metallicity from 9 billion years ago to the present. Here, the ranks, mean, and variance of the SFR are computed on the subset of data from 12 billion years ago until 3 billion years ago. As the average SFR evolves over time, the mean computed on this subset may differ from the mean SFR across the entire duration of 12 billion years. Therefore, the correlation function of the normalized ranks ($\rho^s_{t'} |_{n}(\tau=3~\rm{Gyr})[\rm{SFR}, Z^{\star}_{\rm{O}}]$) does not have to be equal to the actual Spearman correlation ($\rho^s_{t} |_{n}(\tau=3~\rm{Gyr})[\rm{SFR}, Z^{\star}_{\rm{O}}]$). The corresponding sample mean values of these two quantities of time correlations in each of the halo populations are defined as follows:
\begin{align}
\label{eq:def-time-correl}
\rho^s_{t}(\tau)[\rm{SFR}, Z^{\star}_{\rm{O}}] = \avg{ \rho^s_{t} |_{n}(\tau)[\rm{SFR}, Z^{\star}_{\rm{O}}] }_n, \\ 
\rho^s_{t'}(\tau)[\rm{SFR}, Z^{\star}_{\rm{O}}] = \avg{ \rho^s_{t'} |_{n}(\tau)[\rm{SFR}, Z^{\star}_{\rm{O}}] }_n.
\end{align}

\begin{figure}[htbp]
\centering
\includegraphics[width=.49\linewidth]{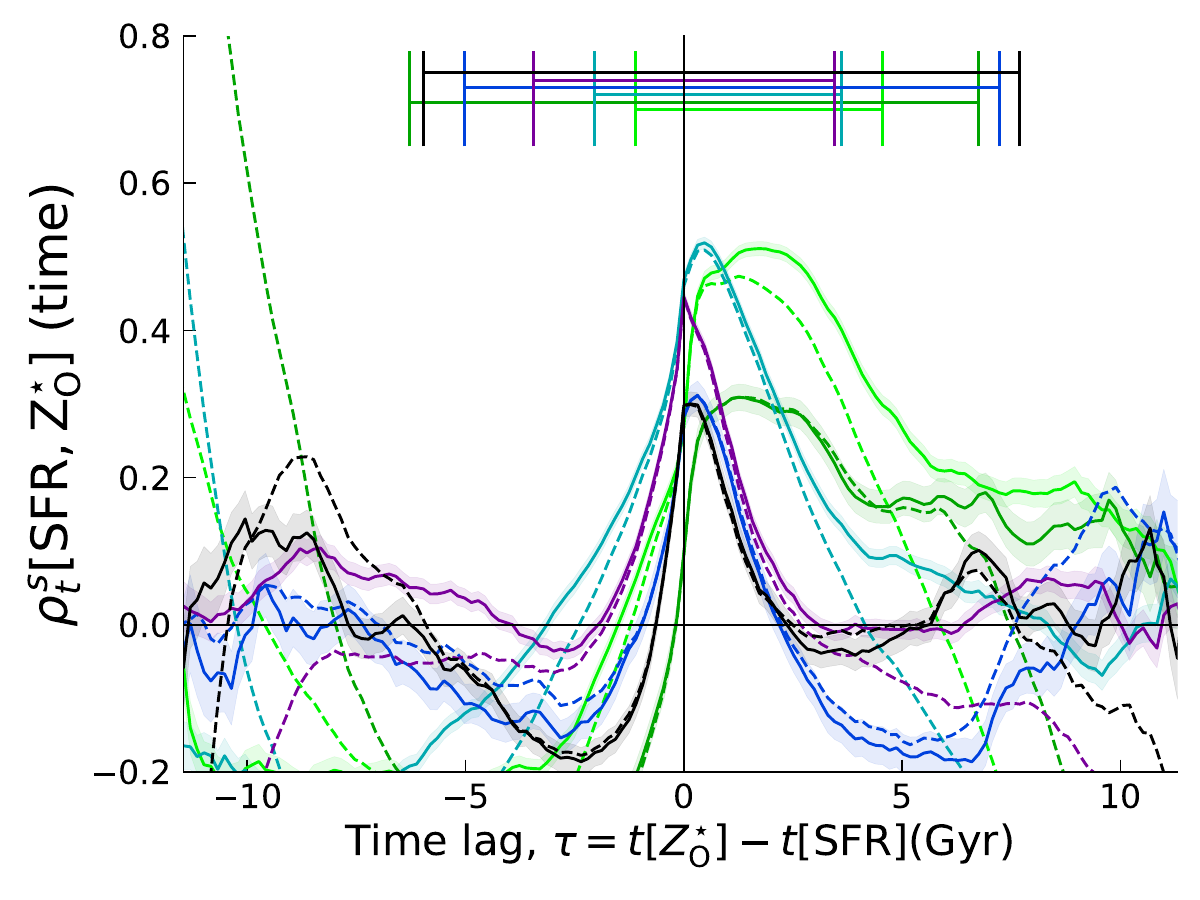}
\includegraphics[width=.49\linewidth]{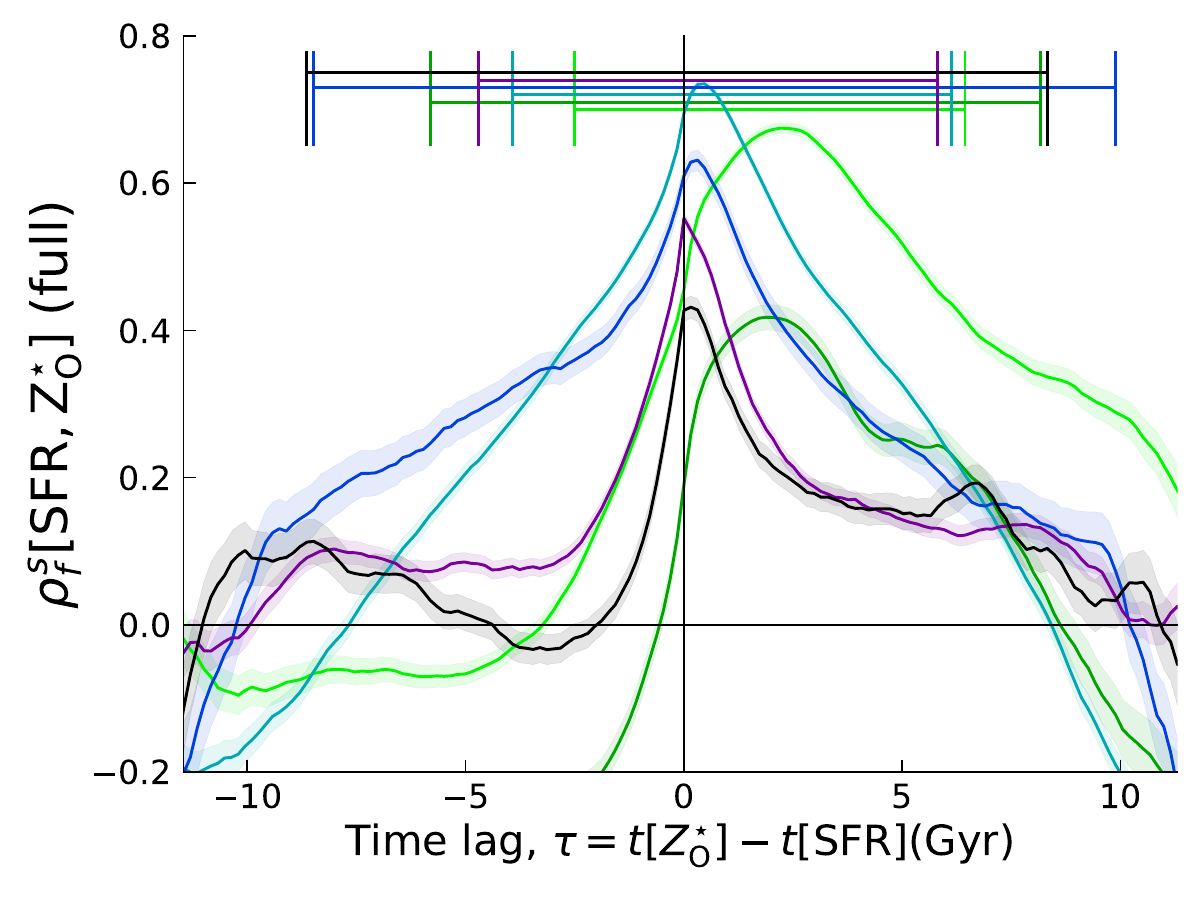}
\caption{Spearman rank correlation coefficients, $\rho^s_t$ (left panel) and $\rho^s_f$ (right panel) as a function of the time lag between SFR and $Z^{\star}_{\rm{O}}$ for haloes selected by their final masses at redshift $z\sim 0$. Additionally, in the left panel the dashed lines indicate cross-correlation $\rho^s_{t'}$ as defined in the main text. Colour coding follows \figref{fig:evolution-hal-gal-props}. The horizontal lines indicate the time lag range that we interpret physically.}
\label{fig:dynam-correl-sfr-ZOsfr-timeshift-func-all}
\end{figure}

These time correlations ($\rho^s_t ~\&~ \rho^s_{t'}$) are shown by the solid and the dashed curves in the left panel of \figref{fig:dynam-correl-sfr-ZOsfr-timeshift-func-all} in each of our halo populations. While they agree at $\tau=0$, they deviate significantly as the time lag becomes larger than the timescales over which the local mean of the quantities evolve. In this work, we will only focus on the time lags where these two quantities are similar, as marked in the \figref{fig:dynam-correl-sfr-ZOsfr-timeshift-func-all}. In that case, we will consider the Spearman correlation as a function of time $\rho^s_t$; It takes on a value between $-1$ and $+1$, similar to the standard Spearman correlation coefficient. 

Notice that the positive time correlation $\rho^s_t>0$ is predominantly seen only with positive time lags $\tau>0$ even at the intermediate halo masses (see blue curves in \figref{fig:dynam-correl-sfr-ZOsfr-timeshift-func-all}). In general, at all halo masses considered, the SFR is correlated with the then future metallicity over a duration of atleast $2~\rm{Gyr}$. In addition, the time correlation also shows some more interesting information. It shows a significant anti-correlation between SFR and the \textbf{past} metallicity ($Z^{\star}_{\rm{O}}$) in most of our halo populations. This can again be interpreted in terms of lowering metallicity by the gas inflow from CGM, which also feeds the future star formation activity. 

The time correlation $\rho^s_t$ focuses on the temporal variations, but it doesn't consider the sample variation in the halo population. We define the full correlation, $\rho^s_f(\tau)$ by simultaneously correlating both across the sample of haloes and over their evolutionary tracks as follows:
\begin{align}
\rho^s_f (\tau)[\rm{SFR}, Z^{\star}_{\rm{O}}] &= \rho^s [ ~\rm{SFR}(n,t), Z^{\star}_{\rm{O}}(n,t+\tau) ~] \quad \text{where } n \in [1,N] ~\&~ t, t+\tau \in [t_1,t_T].
\end{align}
When computed at distinct time steps ($\tau = m \Delta t$) for integer $m=0, \pm1, \ldots, \pm T$, we have,
\begin{align}
\label{eq:def-full-corr}
\rho^s_f (\tau = m \Delta t) &= 
\begin{cases}
\rho^s [ ~\rm{SFR}(\{1 \ldots N\}, ~\{t_1 \ldots t_{T-m}\}), ~Z^{\star}_{\rm{O}}(\{1 \ldots N\}, ~\{t_{1+m} \ldots t_T\}) ~]  \qquad \text{if } m \geq 0 \\
\rho^s [ ~\rm{SFR}(\{1 \ldots N\}, ~\{t_{-m+1} \ldots t_T\}), ~Z^{\star}_{\rm{O}}(\{1 \ldots N\}, ~\{t_1 \ldots t_{T+m}\}) ~] \qquad \text{if } m < 0
\end{cases}.
\end{align}
This considers $(T-m) \times N$ values of both SFR and $Z^{\star}_{\rm{O}}$ at each time lag $\tau = m \Delta t$; Suppose the ranks are denoted as $R^m_{\rm{SFR}}(n,t_i)$ and $R^m_{Z^{\star}_{\rm{O}}}(n,t_i)$, then for a positive value of $m$,
\begin{align}
\rho^s_f (\tau = m \Delta t) &= \frac{ \sum_{i=1}^{T-m} \sum_{n=1}^{N} R^m_{\rm{SFR}}(n,t_i) \times  R^m_{Z^{\star}_{\rm{O}}}(n,t_j) - \mu_{n,t} [ R_{\rm{SFR}}(n,t_i) ]  \times \mu_{n,t} [R_{Z^{\star}_{\rm{O}}}(n,t_j)] }
{ N(T-m) \times \sigma_{n,t} [ R_{\rm{SFR}}(n,t_i) ]  \times \sigma_{n,t} [R_{Z^{\star}_{\rm{O}}}(n,t_j)] }
\end{align}
where the mean ($\mu_{n,t}$), standard deviation ($\sigma_{n,t}$) and the ranks are computed over the $(T-m) \times N$ values,
\begin{align}
R^m_{\rm{SFR}}(\{1 \ldots N\}, ~\{t_1 \ldots t_{T-m}\}) &= R[ ~\rm{SFR}(\{1 \ldots N\}, ~\{t_1 \ldots t_{T-m}\}) ~]\\
R^m_{Z^{\star}_{\rm{O}}}(\{1 \ldots N\}, ~\{t_{1+m} \ldots t_T\}) &= R[ ~Z^{\star}_{\rm{O}}(\{1 \ldots N\}, ~\{t_{1+m} \ldots t_T\}) ~]
\end{align}
This full 2D Spearman correlation coefficient $\rho^s_f[\rm{SFR},Z^{\star}_{\rm{O}}]$ is presented in the right panel of the \figref{fig:dynam-correl-sfr-ZOsfr-timeshift-func-all}.
Once again, we only consider the range of time-lags marked in the figure at each of the halo masses. These are obtained by comparing $\rho^s_f(\tau)$ with the cross-correlation of normalized ranks computed on the entire duration and across all haloes in the population. We find that the full correlation $\rho^s_f$ tends to be stronger than the time correlation $\rho^s_t$, while also including the correlations in temporal variations unlike $\rho^s_h$ defined in \eqref{eq:def-hal-correl}.

\subsubsection{Time lag analysis}
\label{sec:time-lag-analysis}
Spearman correlations ($\rho^s_h, \rho^s_t, ~\&~ \rho^s_f$) as a function time lag ($\tau$) as shown in figures \ref{fig:dynam-correl-sfr-ZOsfr-timeshift-func-halhal} and \ref{fig:dynam-correl-sfr-ZOsfr-timeshift-func-all} depict a qualitative picture of the sequences of galactic processes. In this section, we quantify the typical timescales associated with those processes. In this regard, we compute the mean time lag between SFR and metallicity weighed by their correlation strengths. This is done by considering the function of Spearman correlation coefficients in two regions of time lags: the correlation era $\rho_{+}$ and the anti-correlation era $\rho_{-}$.
\begin{gather}
\rho_{+}(\tau) = 
\begin{cases} 
\rho^s_f(\tau) & \text{if } \rho^s_f(\tau) > 0 \\ 
0 & \text{else} 
\end{cases}, \qquad
\rho_{-}(\tau) = 
\begin{cases} 
-\rho^s_f(\tau) & \text{if } \rho^s_f(\tau) < 0 \\
0 & \text{else} 
\end{cases}
\end{gather}
Then, for both the correlation and anti-correlation, we obtain the correlation weighted mean time lag ($\tau_{\pm}$), the mean correlation strength $A_{\pm}$ in that era and the time duration $\Delta\tau_{\pm}$
over which the correlation is significant.
\begin{gather}
\Delta\tau_{+} = \int_{\rho(\tau)>0} w d\tau , \qquad \Delta\tau_{-} = \int_{\rho<0} w d\tau   \\
A_{+} = \int w(\tau) \rho_{+}(\tau) d\tau / \Delta\tau_{+}, \qquad A_{-} = \int w(\tau) \rho_{-}(\tau) d\tau / \Delta\tau_{-} \\
\tau_{+} = \frac{1}{A_{+} \Delta\tau_{+}} \int \tau w(\tau) \rho_{+}(\tau) d\tau , \qquad \tau_{-} = \frac{1}{A_{-}\Delta\tau_{-}} \int \tau w(\tau) \rho_{-}(\tau) d\tau     
\end{gather}
Here, the weight function $w$ is taken to be the fraction of the entire data between $z \sim 5$ and $z \sim 0$ involved in computing the correlation coefficient at a given time lag. In general, the values of $\tau_{\pm}$ are physical only when the corresponding correlation strengths $A_{\pm}$ and the durations $\Delta \tau_{\pm}$ are both significant. Also note that when correlating between two quantities say $X$ and $Y$ a positive time lag $\tau_{+}[X,Y]$ (or $\tau_{-}[X,Y]$) indicate that the positive (or negative) correlation is seen with $Y$ lagging behind $X$. Hence the mean time lags are anti-symmetric with respect to the quantities, $\tau_{\pm}[X,Y]= - \tau_{\pm}[Y,X]$; However, the strengths and durations of correlations are symmetric, $A_{\pm}[X,Y]= A_{\pm}[Y,X]$ and $\Delta \tau_{\pm}[X,Y]= \Delta \tau_{\pm}[Y,X]$.

\begin{figure}[htbp]
\centering
\conditionalincludegraphics[width=.49\linewidth]{plots/dynam_relxn/shift_betw_SFR-Z(O)_SFreg_fullcorr.pdf}{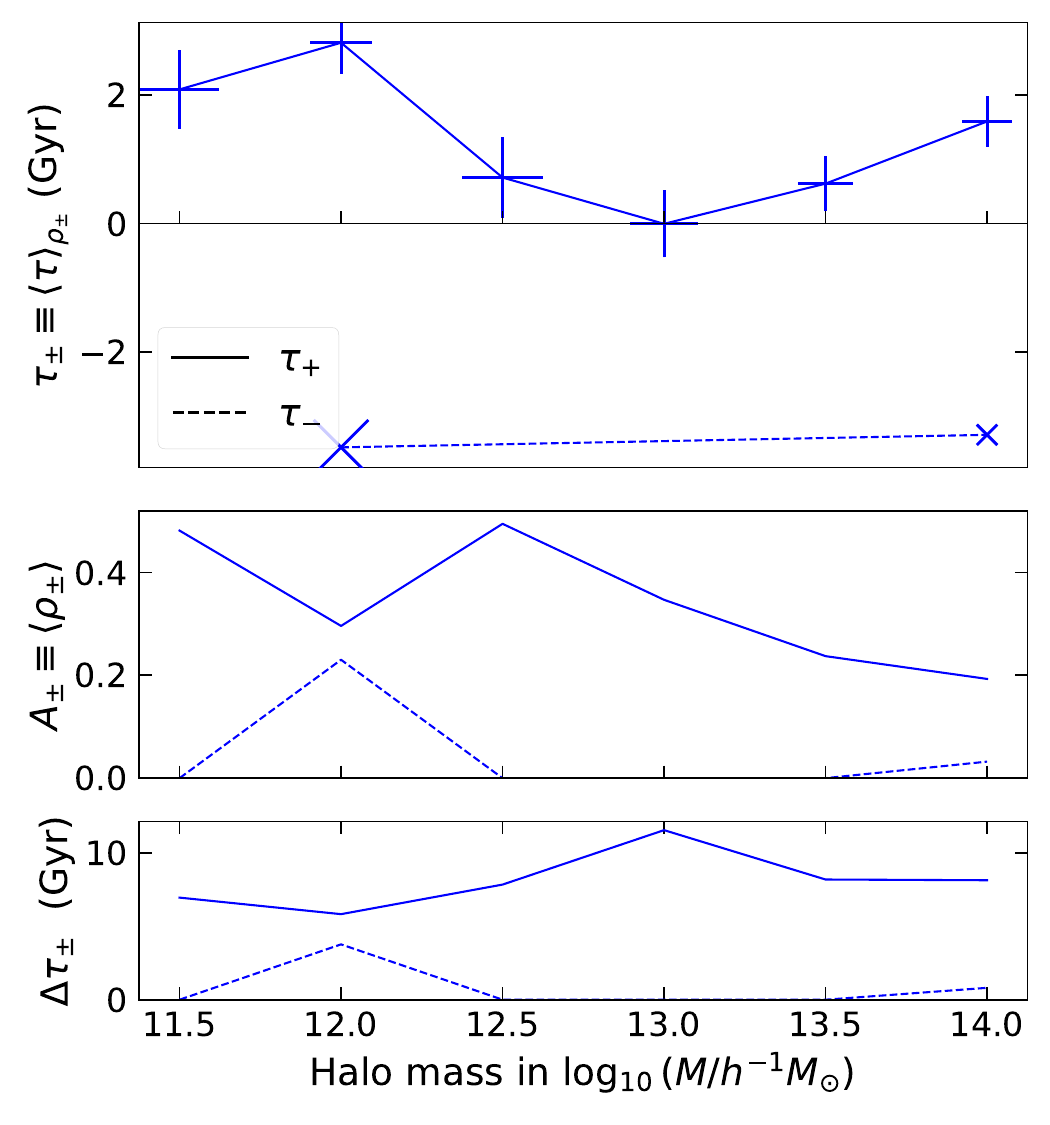}
\caption{In the top panel, correlation weighted mean time lags $\tau_{\pm}$ between SFR and $Z^{\star}_{\rm{O}}$ in the six halo populations selected by their final masses at redshift $z\sim 0$ are shown. The corresponding strengths $A_{\pm}$ and the durations $\Delta \tau_{\pm}$ of the correlation are shown in the middle and bottom panels, respectively. These strengths are also depicted in the upper panel by the marker size.}
\label{fig:dynam-correl-sfr-ZOsfr-timeshift-all}
\end{figure}

The mean time lags ($\tau_{\pm}$), the effective strengths ($A_{\pm}$) and the durations ($\Delta\tau_{\pm}$) of the correlation between the SFR and metallicity $(Z^{\star}_{\rm{O}})$ are shown in the \figref{fig:dynam-correl-sfr-ZOsfr-timeshift-all} for all our halo populations. 
The positive correlation between SFR and future $Z^{\star}_{\rm{O}}$ is depicted by positive values of $\tau_{+}[\rm{SFR}, Z^{\star}_{\rm{O}}]$. At low masses $(<10^{12.5}\Mh)$ this time lag is as large as $\tau_{+}=\sim 2 ~\rm{Gyr}$. At higher masses, though, the time lags are smaller, along with relatively lower correlation strengths. This is likely due to the strong AGN feedback redistributing the metals from the star-forming regions. In addition, the negative correlation identified qualitatively in the previous section can also be seen in the \figref{fig:dynam-correl-sfr-ZOsfr-timeshift-all} with large negative time-lags ($\tau_{-} \sim -3~ \rm{Gyr}$) in some halo populations (shown by dashed curves). The values of $\tau_{-}$ are missing for some of the halo masses, where the correlation $\rho^s_f(\tau)$ is not significantly negative within the range of time lags considered.

\subsection{Interplay of halo relaxation with galactic processes}
\label{sec:main-res-halgal-relxn}
In this section, we study the correlations between the relaxation offset and various other properties of the halo/galaxy. To quantify the strength of the relaxation offset we define $Q_0 \equiv -q_0$; similarly we also define $Q_0^{\rm{in}}\equiv -q_0^{\rm{in}} \quad \& \quad  Q_0^{\rm{out}}\equiv -q_0^{\rm{out}}$ to characterize the strength of the relaxation offset in the inner and the outer regions of the halo respectively. Following the procedure described in the previous section, we primarily focus on the correlation-weighted time lags and their effective correlation strengths in this analysis.

\begin{figure}[htbp]
\centering
\includegraphics[width=.69\linewidth]{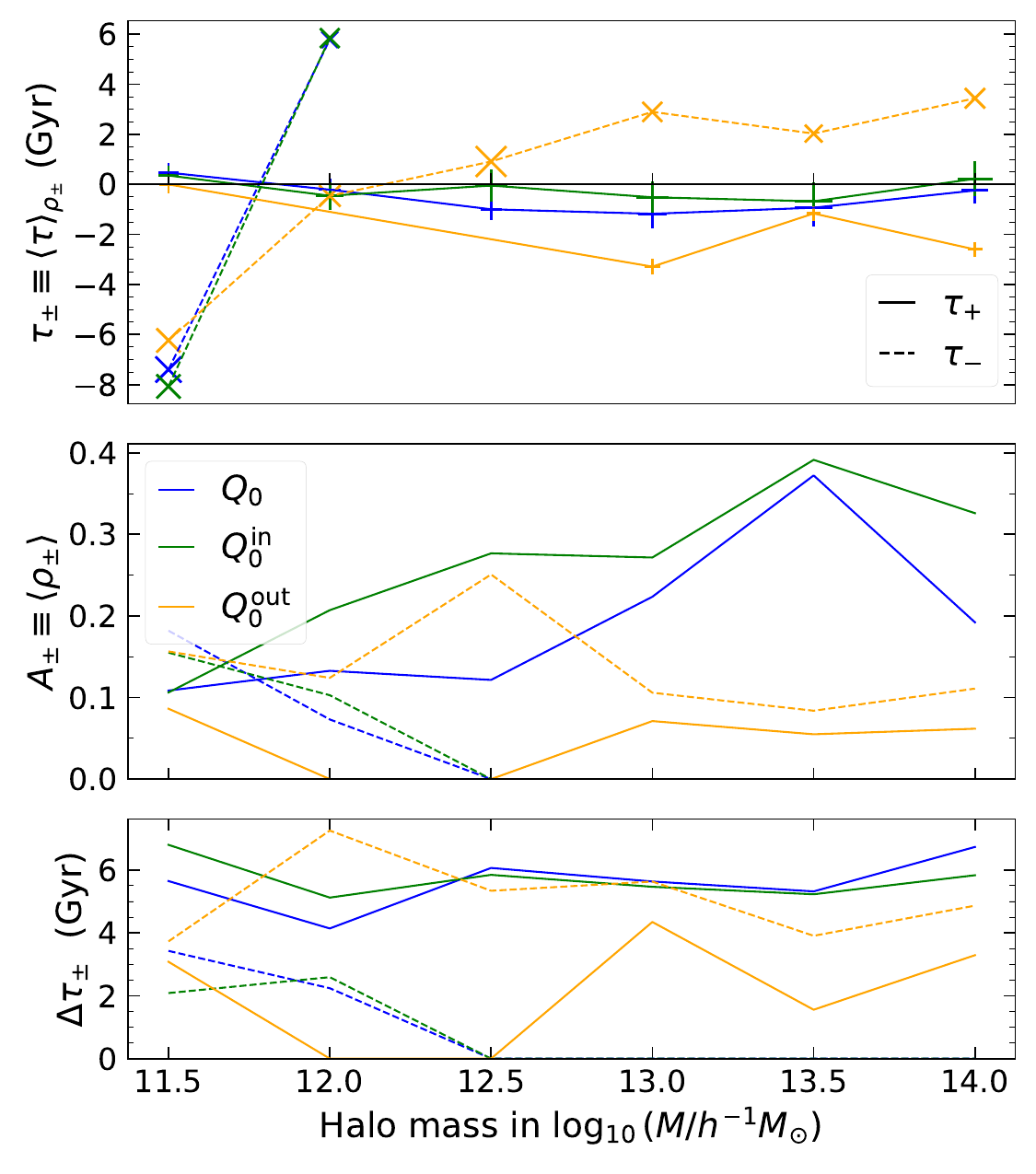}
\caption{In the top panel, correlation weighted mean time lags ($\tau_{\pm}$) between relaxation offset strengths ($Q_0, Q_0^{\rm{in}}, Q_0^{\rm{out}}$) against the star formation rate (SFR) in the six halo populations selected by their final masses at redshift $z\sim 0$ are shown. The corresponding strengths ($A_{\pm}$) and the durations ($\Delta \tau_{\pm}$) of the correlation are shown in the middle and bottom panels, respectively. These strengths are also depicted in upper panel by the size of the markers.}
\label{fig:dynam-correl-q0-SFR-timeshift-func}
\end{figure}

Let us start by studying the correlations between relaxation offset strengths and the SFR shown in the \figref{fig:dynam-correl-q0-SFR-timeshift-func}. In each case, a positive time lag of positive or negative correlation would mean that the corresponding correlation is seen with SFR lagging behind the offset strength; On the other hand, a negative time lag would indicate that the offset strengths lag behind the SFR with that correlation. We find that at all masses $Q_0$ shows a significant correlation with SFR over a long duration of $\Delta \tau_{+}\sim 6$ Gyr with $\tau_{+}$ close to zero. This is more prominent at high masses with effective correlation strength $A_{+}$ as high as $0.35$ at $10^{13.5} \Mh$. This is consistent with our previous work \citep{2023Velmani&Paranjape} where we have found that the relaxation offset to be stronger among haloes with higher specific star formation rate at redshift zero. In addition, we also see that at all masses except $10^{11.5}\Mh$, the mean time lag of positive correlation is negative with a magnitude as large as 1 Gyr in intermediate-mass haloes. This suggests that the higher star formation activity is followed by an enhanced relaxation offset. This means the past SFR is a better predictor of the current relaxation offset.

Also notice that at all masses, $Q_0^{\rm{in}}$ shows a stronger correlation and with a higher time lag $\tau_{+}[Q_0^{\rm{in}},\rm{SFR}] > \tau_{+}[Q_0,\rm{SFR}]$. On the other hand, $Q_0^{\rm{out}}$ shows a significantly weaker correlation but with time lags $\tau_{+}[Q_0^{\rm{out}},\rm{SFR}] < \tau_{+}[Q_0,\rm{SFR}]$. This suggests that the astrophysical processes associated with star formation activity produce relaxation offset starting from the inner to the outer halo. The outer halo shows this positive correlation only for a shorter duration while negative correlation are stronger and seen for a longer duration ($\Delta \tau_{-}\sim 6 ~\rm{Gyr}$). With $\tau_{-}[Q_0^{\rm{out}},\rm{SFR}]>0$, this negative correlation with SFR is mostly seen over past relaxation offset in the outer halo.

\begin{figure}[htbp]
\centering
\conditionalincludegraphics[width=.69\linewidth]{plots/dynam_relxn/shift_betw_multi-Z(O)_SFreg_fullcorr.pdf}{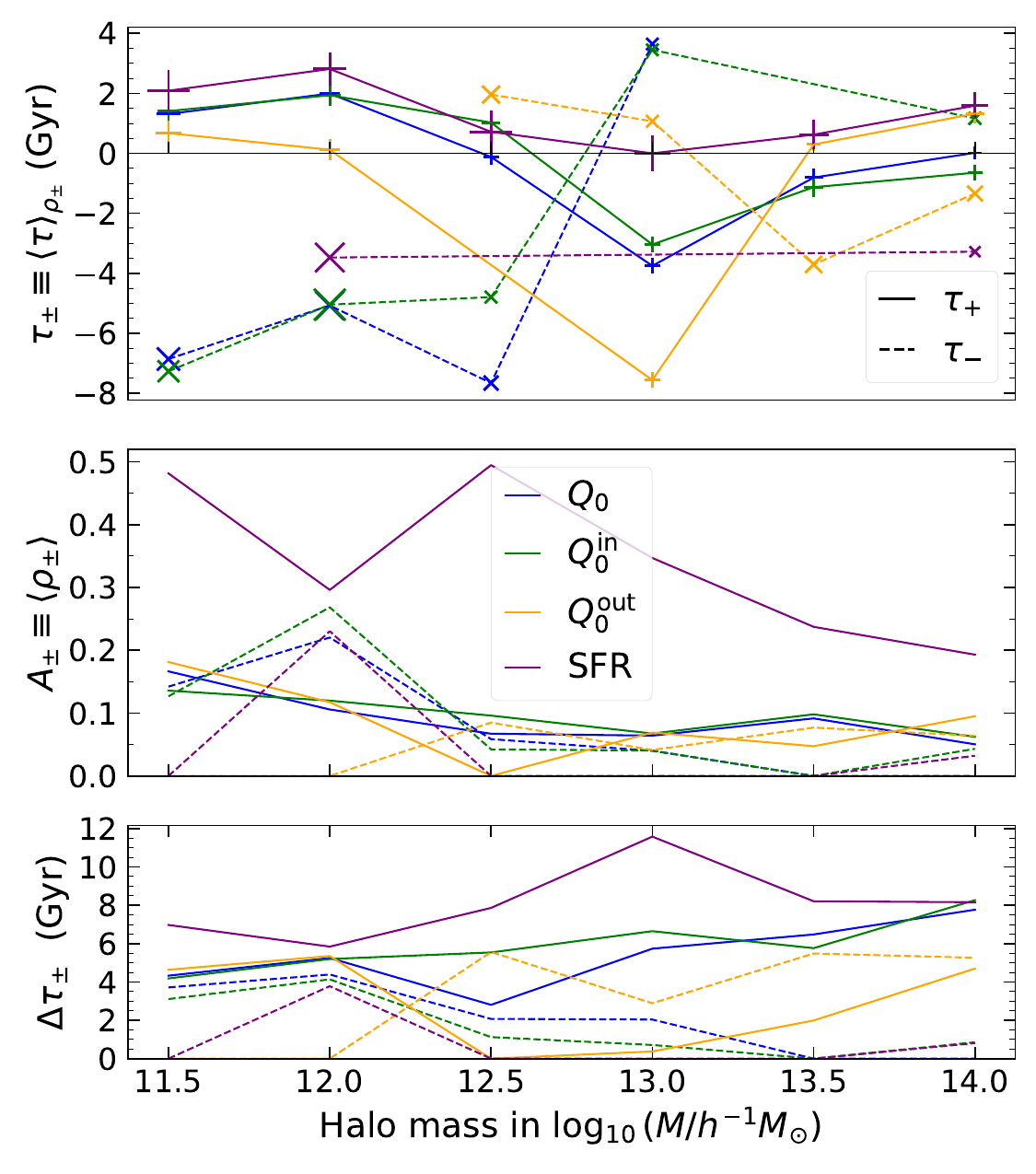}
\caption{In the top panel, correlation weighted mean time lags ($\tau_{\pm}$) between relaxation offset strengths ($Q_0, Q_0^{\rm{in}}, Q_0^{\rm{out}}$) and SFR against the metallicity ($Z^{\star}_{\rm{O}}$) in the six halo populations selected by their final masses at redshift $z\sim 0$ are shown. The corresponding strengths ($A_{\pm}$) and the durations ($\Delta \tau_{\pm}$) of the correlation are shown in the middle and bottom panels, respectively. These strengths are also depicted in upper panel by the size of the markers.}
\label{fig:dynam-correl-q0-ZOsfr-timeshift-func}
\end{figure}

Moreover, the above result suggests that the effect of star formation activity on the relaxation offset is more immediate on the inner halo compared to the outer halo. This is consistent with the argument that the feedback from massive, short-lived stars that follow the star formation activities is causing the offset in the relaxation.  However, this also means that a measure of the recent feedback processes should correlate even more strongly with the relaxation offset strength. To test this, we have studied the correlations between the relaxation offset strengths and metallicity of oxygen in the star-forming regions ($Z^{\star}_{\rm{O}}$); this is shown in \figref{fig:dynam-correl-q0-ZOsfr-timeshift-func}

We find that the relaxation offset strength does show a positive correlation with $Z^{\star}_{\rm{O}}$, albeit weaker than the correlation found with SFR. 
With the exception of the highest halo masses, the positive correlation is first seen in the inner halo followed by the outer halo. This is depicted by 
\begin{align}
\tau_{+}[Q_0^{\rm{in}},Z^{\star}_{\rm{O}}] > \tau_{+}[Q_0^{\rm{out}},Z^{\star}_{\rm{O}}] > \tau_{+}[Q_0,Z^{\star}_{\rm{O}}]
\end{align}
Also, the correlations are relatively weaker for high-mass haloes where the correlation between SFR and $Z^{\star}_{\rm{O}}$ is also weaker. In $10^{13} \Mh$ haloes, there is a significant correlation with SFR ($A_{+}[\rm{SFR},Z^{\star}_{\rm{O}}] \sim 0.35$), but interestingly $Q_0$ correlates with much earlier $Z^{\star}_{\rm{O}}$ of over 4 Gyr in the past, while it correlates with SFR at just 1 Gyr in the past on average.
This may be because the feedback alone does not mediate the relaxation offset. At least we can say that the metallicity is not strongly associated with the feedback processes related to star formation activity involved in producing relaxation offset.

\begin{figure}[htbp]
\centering
\conditionalincludegraphics[width=.69\linewidth]{plots/dynam_relxn/shift_betw_multi-dM(O)_FoF_fullcorr.pdf}{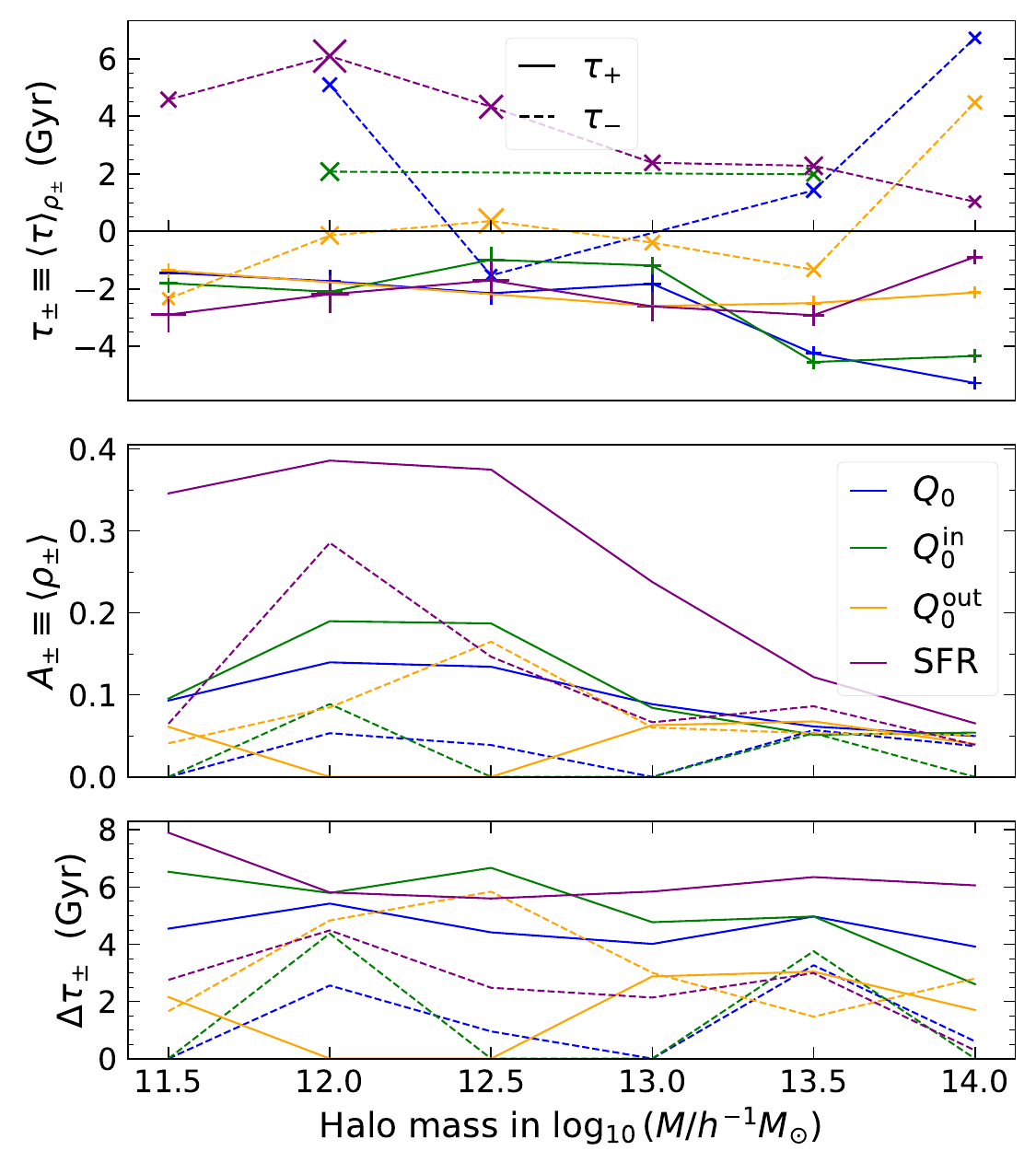}
\caption{In the top panel, correlation weighted mean time lags ($\tau_{\pm}$) between relaxation offset strengths ($Q_0, Q_0^{\rm{in}}, Q_0^{\rm{out}}$) and SFR against the rise in the oxygen mass in the gas ($M_{\rm{O}}'$) in the six halo populations selected by their final masses at redshift $z\sim 0$ are shown. The corresponding strengths ($A_{\pm}$) and the durations ($\Delta \tau_{\pm}$) of the correlation are shown in the middle and bottom panels, respectively. These strengths are also depicted in upper panel by the size of the markers.}
\label{fig:dynam-correl-q0-dMOFoF-timeshift-func}
\end{figure}

Since the metallicity in the star forming regions is also affected by the inflowing gas and redistribution by AGN feedbacks, we also study correlations with the overall rise in metals. In particular, let us now focus on the correlations between relaxation offset strengths and the rise in total oxygen mass in the gas content of the entire halo ($M_{\rm{O}}'$).  While $Z^{\star}_{\rm{O}}$ is an integrated quantity, $M_{\rm{O}}'$ tracks instantaneous increase in the metals. However, $M_{\rm{O}}'$ also shows only a weaker correlation with the relaxation offset compared to correlations with SFR (see \figref{fig:dynam-correl-q0-dMOFoF-timeshift-func}). We find that the loss of metals in the gas to the newly formed stars dominates over the metal enrichment by feedback. Hence SFR is negatively correlated with the future $M_{\rm{O}}'$ 
depicted by positive values of $\tau_{-}[\rm{SFR},M_{\rm{O}}']$.

\begin{figure}[htbp]
\centering
\includegraphics[width=.69\linewidth]{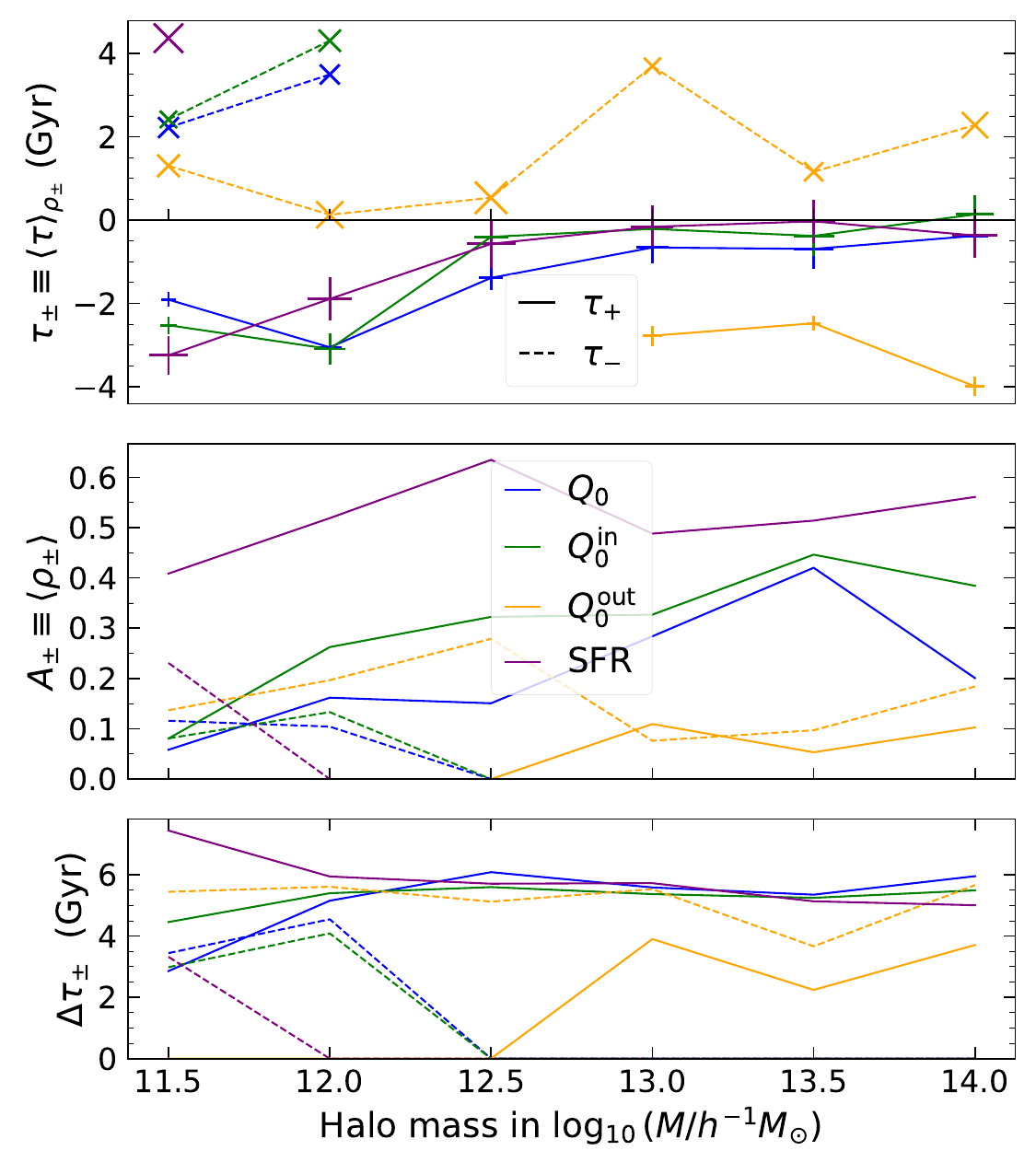}
\caption{In the top panel, correlation weighted mean time lags ($\tau_{\pm}$) between relaxation offset strengths ($Q_0, Q_0^{\rm{in}}, Q_0^{\rm{out}}$) and SFR against the mass in wind within the halo ($M_{\rm{Wind}}$) in the six halo populations selected by their final masses at redshift $z\sim 0$ are shown. The corresponding strengths ($A_{\pm}$) and the durations ($\Delta \tau_{\pm}$) of the correlation are shown in the middle and bottom panels, respectively. These strengths are also depicted in upper panel by the size of the markers.}
\label{fig:dynam-correl-q0-Mwind-timeshift-func}
\end{figure}

Finally, we have also studied the correlations with the wind mass $M_{\rm{Wind}}$; this is presented in \figref{fig:dynam-correl-q0-Mwind-timeshift-func}. We find that the wind mass correlates with relaxation offset strengths in the future, affecting the inner halo earlier than the outer halo. We also find that the relaxation offset strength is correlated (although weakly) with the wind mass at a much later time in the future in low-mass haloes compared to high-mass haloes. We do see a similar trend in the correlation between wind mass and SFR; wind mass positively correlates with the star formation activity at a later time for lower mass haloes. However, we note that for the $10^{11.5}\Mh$ haloes, the wind mass correlates with the star-forming activity nearly 3.5 Gyr later. On the other hand, it correlates with the relaxation offset strength at just 2 Gyr in the future. Also, this correlation between wind mass and the relaxation offset is slightly stronger than the correlation we saw with SFR. This suggests that the wind mass is likely more closely associated with the relaxation offset than the star formation rate itself.

\section{Conclusion}
\label{sec:conclusion}

This study explored the dynamical evolution of dark matter's response to galaxies within populations of haloes simulated using the IllustrisTNG cosmological volumes. By constructing a detailed population of haloes and tracing their evolutionary tracks, we characterized the relaxation response of these haloes. This was performed by comparing haloes from hydrodynamical simulations, which include subgrid prescriptions for various astrophysical processes, with corresponding haloes from gravity-only simulations. Using a catalogue of evolving matched haloes, we examined the correlation between relaxation quantities and other halo/galaxy properties to elucidate their roles in mediating the relaxation response.

Firstly, we find that the radially-dependent linear relaxation relation model proposed in our previous work is applicable even at earlier redshifts, at least from redshift $z=5$. In this work, we have primarily studied the offset parameter $q_0$ in the relaxation relation that characterizes the amount of relaxation of the dark matter shells with no change in the enclosed mass. In a given population of haloes selected by their final mass, this offset is on average stronger during the peak star formation among those haloes. 
Our findings reveal that star formation activity significantly influences the offset in the halo relaxation response  over the entire evolutionary history of the haloes. While this connection with SFR is immediate on the relaxation in the inner haloes, it is seen 2 to 3 billion years later on average in the outer regions of both Milky Way scale haloes and halo groups. %
We also found that simple tracers of the stellar feedback processes through metal content only show a weaker connection with the relaxation than SFR itself. However, the wind accumulated from various feedback processes did have a stronger connection with the relaxation.%

These insights enhance our understanding of the mechanisms driving halo relaxation and contribute to the development of more accurate models of halo profiles in baryonification procedures. 
For example, the knowledge of time lags can, in princple, allow modelling the observed dark matter distribution in large surveys such as Euclid with fewer parameters by exploiting correlations across different redshift bins.
And in semi-analytic galaxy formation models, this %
may allow simple time-dependent transformation procedures to incorporate the dynamical evolution of the host dark halo with galaxy evolution, which is typically ignored. In the future, the relaxation response of dark matter haloes can also serve as a probe into the evolutionary history of the galaxies they host.

\section*{Data availability}
No new data were generated during the course of this research.

\acknowledgments
The research of AP is supported by the Associates scheme of ICTP, Trieste.
We gratefully acknowledge the use of high performance computing facilities at IUCAA (\url{http://hpc.iucaa.in}). This work made extensive use of the open source computing packages NumPy \citep{vanderwalt-numpy},\footnote{\url{http://www.numpy.org}} SciPy \citep{scipy},\footnote{\url{http://www.scipy.org}} Matplotlib \citep{hunter07_matplotlib},\footnote{\url{https://matplotlib.org/}} Pandas \citep[][]{reback2020pandas},\footnote{\url{https://pandas.pydata.org/about/}} Schwimmbad \citep{schwimmbad},\footnote{\url{https://joss.theoj.org/papers/10.21105/joss.00357}} H5py,\footnote{\url{https://www.h5py.org/}} Colossus \citep{colossus},\footnote{\url{http://www.benediktdiemer.com/code/colossus/}}  Jupyter Notebook\footnote{\url{https://jupyter.org}} and Code-OSS.\footnote{\url{https://github.com/microsoft/vscode}}

\bibliography{references,references_new}

\end{document}